\PassOptionsToPackage{hidelinks}{hyperref}
\PassOptionsToPackage{dvipsnames}{xcolor}
\documentclass[pmlr]{jmlr}% new name PMLR (Proceedings of Machine Learning Research)

\usepackage{longtable}% for long tables

\usepackage{booktabs}
\usepackage[load-configurations=version-1]{siunitx} % newer version

\usepackage{times}
\usepackage{soul}
\usepackage[utf8]{inputenc}
\usepackage[T1]{fontenc}
\usepackage[small]{caption}
\usepackage{algorithm}
\usepackage{algorithmic}

\usepackage{enumitem}
\usepackage[para]{footmisc}
\usepackage{xspace}
\usepackage{wrapfig}
\usepackage{tcolorbox}
\usepackage{tikz}
\usepackage{multirow}

\usepackage{verbatim}
\usepackage{makecell}
\usepackage{csquotes}
\usepackage{ifthen}
\usepackage{colortbl}

\usepackage{listings}
\newcommand{\E}{\mathbb{E}}

\jmlrvolume{}          % leave EMPTY — we fill in the volume number
\jmlryear{2026}
\jmlrworkshop{Impactful and Responsible AI Systems for Education}

\title[MRT of AI-Simulated Practice for Therapeutic Skills]{Micro-Randomized Trial of an AI-Simulated Practice Tool for Therapeutic Skills}

\author{\Name{Ryan Louie} \Email{rylouie@cs.stanford.edu}\\
\addr Stanford University
\AND
\Name{Ellen Converse} \Email{ellencon@stanford.edu}\\
\addr Stanford University and Palo Alto University
\AND
\Name{Diyi Yang} \Email{diyiy@stanford.edu}\\
\addr Stanford University
\AND
\Name{Emma Brunskill} \Email{ebrun@cs.stanford.edu}\\
\addr Stanford University
}

\begin{document}

\maketitle

\begin{abstract}
Though LLM-simulated practice may support psychotherapy education, open questions remain about which system features provide the most educational value. We developed an AI-simulated practice tool for use in psychotherapy classrooms that provides speaking-practice with LLM-based patients, followed by a post-practice activity that includes feedback from a fine-tuned LLM and written reflection exercises. We deployed the tool in a graduate psychotherapy course (n=25, 5 weeks) using a micro-randomized trial (MRT)---a method which can estimate the causal excursion effect of an intervention by leveraging longitudinal repeated randomization with participants. Testing four conditions crossing AI feedback (present/absent) with reflection granularity (utterance/session-level), we found surprising interaction effects on student engagement and educational value: utterance-level reflection paired with AI feedback significantly outperformed all conditions. Students valued comparing the AI suggestions to their own rather than passively accepting them. 
% Since education outcomes need to be evaluated
% Our discussion reflects on the importance of evaluating AI models and accompanying learning activities and the promise of repeated-randomization methods like MRTs to support formative, ecologically-valid experiments of AI education research systems.
% experimental paradigms to explore the how design of education activities surrounding AI education tools can impact learning experiences of students in classroom settings.
As the first MRT in a health education setting testing an LLM-simulation system, this work demonstrates that MRTs, a method mostly used in mobile intervention research, offers a viable evaluation paradigm for AI systems deployed in health education settings.
\end{abstract}


\section{Introduction}\label{sec:introduction}
LLM-simulated roleplay practice is an AI education approach to training professional social skills (e.g., interpersonal skills~\cite{lin2024imbue}, and conflict resolution~\cite{rehearsal}), with recent lab studies~\cite{louie2024roleplay,wang2024patient,steenstra2025scaffolding} demonstrating how AI systems can provide realistic practice for training novices for mental health counseling skills. An important challenge when moving these tools from the lab to real-world classrooms is evaluating and improving their pedagogical utility within authentic classroom use. Few studies to date have deployed LLM-simulated roleplay practice in classrooms~\cite{thesen_generative_2025}, and it is uncommon for such studies to evaluate the impact of design choices in real-world educational settings.

Across a two-quarter course sequence, we deployed an AI-augmented practice tool for therapeutic skills which provides utterance-level AI feedback on transcripts of roleplay practice sessions. We found that students were not passive recipients of AI feedback—they critically evaluated suggestions and debated merits of their responses versus AI alternatives. These formative findings during the initial half of our deployment provided valuable insights into the design space of post-practice activities, and motivated a continued experiment (N=25 students, 5-weeks) where we experimentally compared different post-practice activities. These activities guide psychotherapy students to reflect on their own approach and/or the AI's (when AI feedback is available). Reflection prompts are provided at the session-level (broadly about overall approach) or utterance-level (detailed analysis of specific therapeutic responses).

Our experiment employs micro-randomized trials (MRTs)~\cite{klasnja2015microrandomized}—a method commonly used to evaluate mobile health interventions—to AI education settings. MRTs use within-person randomization at each engagement opportunity, gaining statistical power over longer time periods common in education. This enables estimating the marginal causal excursion effect~\cite{qian2021estimating}, which measures the momentary impact of delivering a specific intervention. Students in the AI feedback with utterance-reflection condition reported the greatest learning utility (5.90/7, treatment effect 1.29, $p<0.05$) compared to the no-feedback, session-reflection baseline (4.61/7). This condition also increased reflection engagement nearly nine-fold. Critically, students mentioned integrating AI suggestions rather than passively accepting them, providing evidence that our reflection design, inspired by cognitive forcing functions~\cite{buccinca2021trust} and ~\cite{kolb2014experiential} experiential learning theory, can be used to scaffold critical reflection to support learning with AI feedback.

In the remainder of this paper, we present the CARE platform that was deployed to a graduate psychotherapy course; the micro-randomized trial design for evaluating the impact of post-practice review activities; and detailed results of our micro-randomized trial. Finally, we discuss the implications for how novel experimental methods can be used during formative and continual educational deployments to evaluate and improve the pedagogical utility of AI educational approaches like LLM-simulated training systems.

\begin{figure*}[t]
    \centering
    \includegraphics[width=\linewidth]{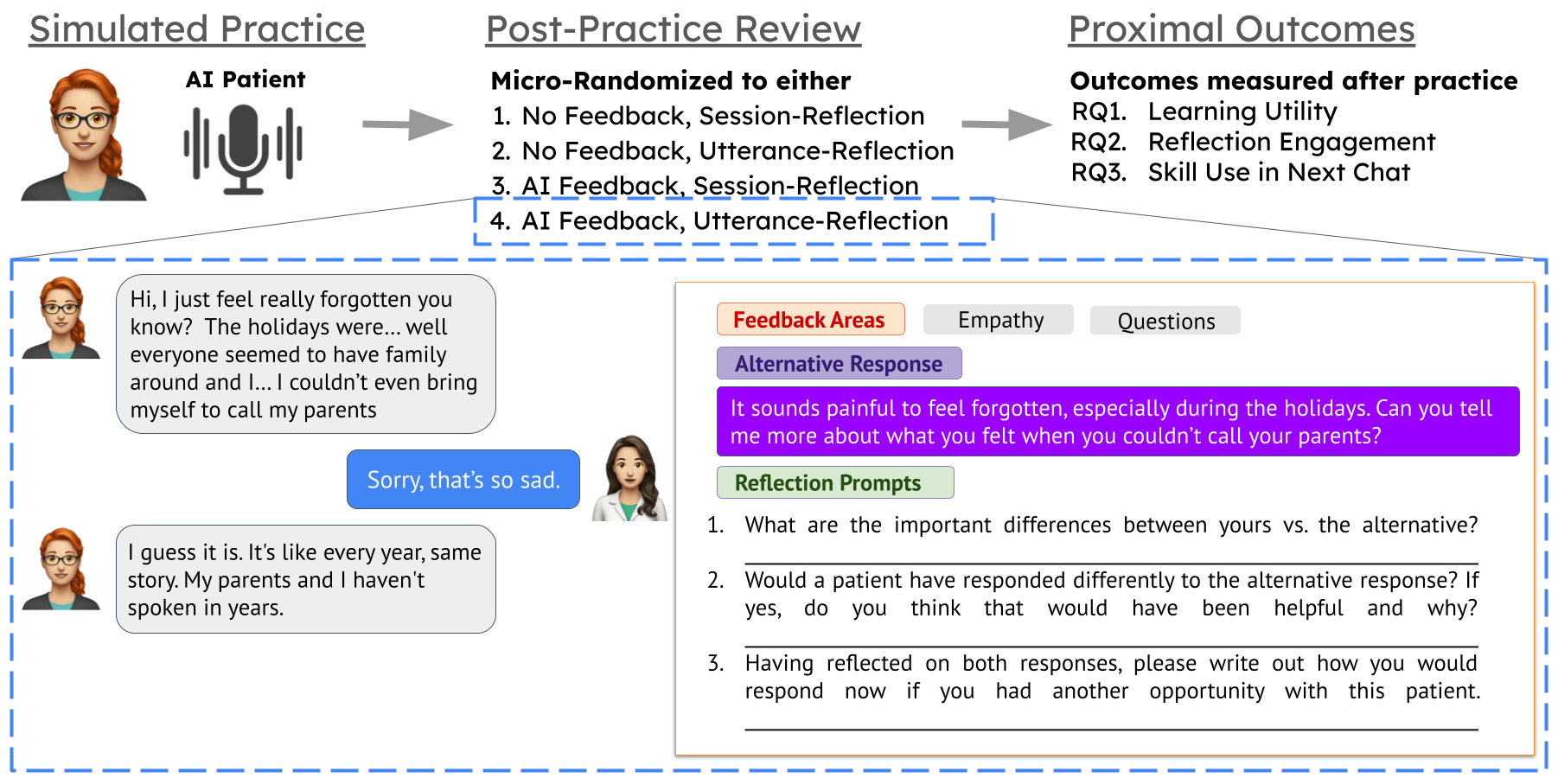}
    \caption{CARE is AI-based psychotherapy training system that provides speaking-practice with LLM-simulated patients, followed by post-practice review activities which vary in AI feedback (present/absent) and granularity of reflection (session-level/utterance-level). To test what type of review activity is most effective, we conducted a micro-randomized trial in a clinical psychology graduate course (5 weeks, 25 students), where students were randomized to one of four categorical treatment arms after AI-simulated practice, and three proximal outcomes were measured after completing the review activity: perceived learning utility (student responses to two Likert-scales items); reflection engagement behaviors (amount of reflection content a student writes), and skill use in the next chat (scored via NLP classifiers).
    % \diyi{I like this figure, but the current look is blurred and I wonder instead of a raw figure, could we do a more visualization mockup of what it looks like, to make it clear? the current highlight seems to be the care interface, but it should be MRT conditions, etc. I also don't think the care interface clearly shows all the options or features we have.}
    %
    }
    \label{fig:fig1}
\end{figure*}

\section{Related Work}\label{sec:related-work}
% \subsection{Towards Scalable Psychotherapy Training}
%

\paragraph{AI for Training Psychotherapy Skills}
Deliberate practice (DP) has become an established framework for developing therapeutic expertise~\cite{nurse2024influence,miller2020or} with empirical evidence supporting its effectiveness~\cite{larsson2025does}. However, widespread implementation is limited by the significant time and expertise required for supervision and personalized feedback. This has led researchers to explore AI-powered systems that automate simulation and feedback~\cite{wang2024patient,hailey,care,steenstra2025scaffolding}. Results from lab studies suggest AI tools can enhance novice therapists' confidence and skill, but questions remain about effectiveness during sustained, education usage~\cite{ajluni2025artificial,yamamoto2024enhancing,wang2024patient}. Our study deployed a roleplay practice tool in a graduate psychotherapy course, to understand what design factors impact the learning experience of students using reviewing their roleplay practice sessions, either with AI-simulated partners or human peers.
% To our knowledge, our system is one of the first to create patient-scenarios matched to a psychotherapy curriculum; and the first to use multimodal-native LLMs to develop emotionally-expressive voice-based patients using a Constitutional AI principle-approach~\cite{louie2024roleplay}---something not afforded by speech-to-text-to-speech approaches.

\paragraph{Experimental Methods for System Deployments}
A general challenge in evaluating systems is balancing ecological validity with understanding causal mechanisms. \cite{klasnja2011evaluate} argues that evaluating proximal outcomes enables rigorous efficacy evaluations within the resource constraints of field studies. Building on this, micro-randomized trials (MRTs) estimate treatment effects by randomizing different intervention versions to the same participants over multiple decision points~\cite{klasnja2015microrandomized}. MRTs have been adopted across physical and mental health interventions, including physical activity promotion~\cite{coppens2024balancing,figueroa2022daily}, substance abuse management~\cite{rabbi2018toward}, and mood management~\cite{arevalo2024effect,zhao2024evaluate}. \cite{cho2021applying} proposed extending MRTs to education, and realizing the potential of mobile interventions for education, \cite{breitwieser2024realizing} implement MRTs in an educational setting. Compared to other within-subject designs such as crossover or round-robin, MRTs offer distinct advantages: each engagement opportunity serves as a decision point, enabling more frequent randomization than fixed time-periods. Unlike crossover designs that assume a stable baseline and require washout periods, treatment effect estimates in MRTs explicitly model participant states as time-varying moderators. To our knowledge, this study is among the first empirical MRTs in health education and the first LLM-based classroom MRT with categorical arms.

% Our work departs from typical MRT experiments in three-ways. First, this study is among the first empirical MRTs in health education. Second, we micro-randomize across four categorical conditions rather than binary treatments, requiring a recently developed causal excursion estimator for categorical treatments~\cite{lin2025micro}, Third, whereas mobile interventions deploy push-based interventions with frequent, system-controlled decision points, our health education setting micro-randomizes at student-initiated moments--requiring tailoring of the causal excursion estimator for time-varying decision points.

\section{AI-Simulated Practice Tool for Psychotherapy Classrooms}\label{sec:system}

Our classroom deployment of CARE, an AI-system for psychotherapy skills training, is a first step at translating previous lab-based findings on LLM-simulated practice and feedback into  real-world educational setting~\cite{louie2025can,louie2024roleplay,chaszczewicz2024multi}.
Our tool was offered as a web-based platform that enables psychotherapy students to practice a {\em voice-based, multi-turn} conversation with an LLM-simulated patient. After a simulated practice session, students can optionally receive LLM-generated feedback on their responses and reflection questions about their practice and/or the feedback they received. 

\textbf{Voice-based LLM-simulated patients.} CARE extends an existing method for defining realistic
LLM-patient behaviors that uses Constitutional AI principles which were elicited from domain-experts during interactive testing~\cite{louie2024roleplay}. Each of CARE's patients integrate expert-validated behavioral principles released by \cite{louie2024roleplay} along with additional principles that were defined during formative testing of the voice-based LLM-patients. In total, 30 voice-based LLM-simulated patients were deployed for the course's 10 week curriculum. The appendix provides more details on the creation of the voice-based LLM-simulated patients and the OpenAI Real-time API implementation of the patients.

\textbf{Post-practice LLM-generated feedback.} CARE integrates an existing method for generating counseling feedback that fine-tunes and self-improves the Llama-2 13B parameter model using an expert-annotated feedback dataset of peer counseling transcripts~\cite{chaszczewicz2024multi}. We choose this fine-tuned model because it generates contextual feedback at multiple-levels, mirroring psychotherapy supervision structure: (1) assess trainee response by highlighting strengths and areas needing improvement across eight microskills (empathy, reflections, questions, validation, suggestions, session management, professionalism, and self-disclosure), (2) generate goal-oriented explanation of the feedback and a suggested alternative.%, enabling trainees to compare their choices against the AI's.

\textbf{Post-practice reflection interfaces.} CARE's utterance-level reflection (Table~\ref{tab:utterance-reflection-questions}) was designed to address evidence that simply presenting AI recommendations leads to superficial processing~\cite{bansal2021does,buccinca2020proxy}. Rather than presenting AI-generated alternatives as recommendations to accept or reject, the interface prompts students to compare their original utterance with the alternative side-by-side. Students then formulate their own revised response through a three-stage Intention-Impact-Revision sequence, compelling deeper cognitive engagement rather than passive acceptance of AI suggestions~\cite{schwartz2016abcs} (see Appendix~\ref{appendix:reflection-theory} for theoretical grounding).

To test the impact of this reflection design, we built three alternative variants: (1) reviewing AI feedback while reflecting on the session holistically; (2) reflecting on a subset of utterances without AI feedback; and (3) reflecting on the session holistically without AI feedback, using two questions covering what was challenging and what the student would revise. These variants allow us compare the impact of variations of AI feedback and reflection granularity together.

\begin{figure*}[t]
    \centering
    \includegraphics[width=\linewidth]{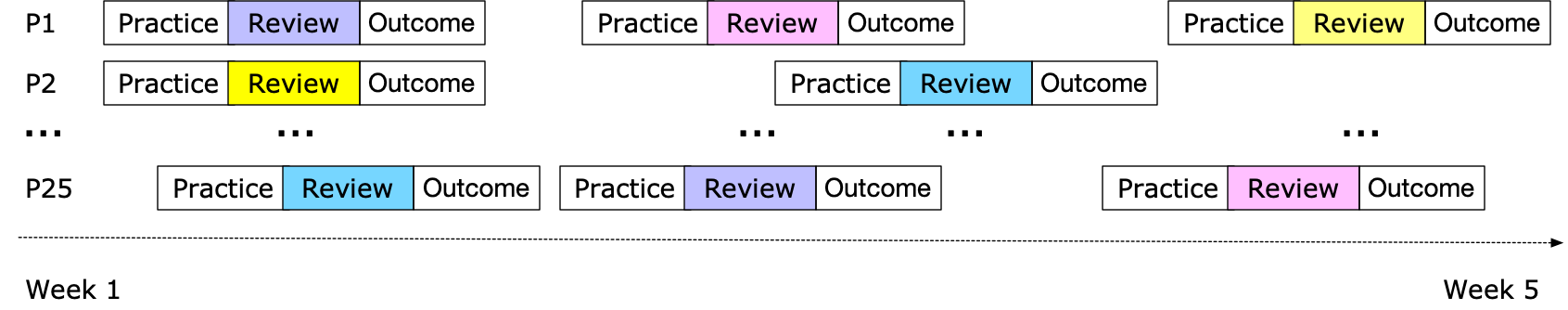}
    \caption{Students (n=25) practiced with AI-simulated patients; after each \emph{student-initiated} session, the system micro-randomized one of four post-practice review activities: \colorbox{yellow!80}{No Feedback, Session Reflection}, \colorbox{Lavender!60}{No Feedback, Utterance Reflection}, \colorbox{ProcessBlue!50}{AI Feedback, Session Reflection}, \colorbox{blue!30}{AI Feedback, Utterance Reflection}. Because learners choose when to practice, decision points (and spacing between them) are irregular; the main text below summarizes measures and analysis.}
    \label{fig:pN-study-randomization}
\end{figure*}

\section{Micro-randomized Trial Design}\label{sec:mrt-design}

To evaluate which post-session review activities best support learning from AI-simulated practice, we ran a Micro-randomized Trial (MRT)~\cite{klasnja2015microrandomized} in a psychotherapy class: treatment is randomized at many time points per participant, not once at enrollment. After each session, the system assigned one of four post-practice conditions with equal probability, independent of history (Figure~\ref{fig:pN-study-randomization}).

\paragraph{Factorial, categorical MRT.} The classroom required a 2$\times$2 over components (AI feedback vs.\ no feedback $\times$ session- vs.\ utterance-level reflection), not a single intervention vs.\ control. We used uniform 25\% randomization to each of four arms at every decision point and report contrasts on proximal outcomes with the \emph{causal excursion effect} (CEE) for categorical treatments~\cite{lin2025micro,boruvka2018assessing}, summarizing \emph{immediate, marginal} effects relative to a reference level. Full potential-outcomes definition, centering, and the weighted least-squares implementation we used are in Appendix~{\ref{appendix:mrt-cee-categorical}}; immediate outcomes (vs.\ delayed $t{+}\Delta$) are standard in MRT practice~\cite{lin2025micro}.

\paragraph{Student-initiated (pull) sessions.} In contrast to many mobile-health MRTs with clocked or push notifications, each decision point occurred when a student \emph{started} a practice session, so the number and timing of randomizations differed by student. Inter-session gaps in our data were empirically irregular; we analyzed all person-decision points (Appendix~{\ref{appendix:mrt-pull-irregular}}), so effects average over the engagement patterns that arose naturally, rather than a fixed weekly pull schedule. A software issue in session-reflection conditions meant some practices did not display the prompt; for those times we set availability $I_{i,t}=0$ in estimation (Appendix~{\ref{appendix:mrt-availability-sessions}}; see Table~{\ref{tab:cee-rq1-noavail}} for a sensitivity that sets $I_t{=}1$ everywhere).

\paragraph{RQs.} MRTs compare intervention \emph{components} rather than confirming a single package. We ask: \textit{RQ1} perceived learning from review; \textit{RQ2} reflection behavior; \textit{RQ3} skill use in the next session; and \textit{RQ4} course-end views of the review variants.

\paragraph{Measures.} \textbf{Perceived learning utility} (RQ1) was the mean of two 7-point items administered immediately after review (item wording, Appendix~{\ref{appendix:mrt-measures}}). \textbf{Reflection engagement} (RQ2) was the log of total reflection text length, normalized by the number of prompts in that review type. \textbf{Skill use} (RQ3) used fine-tuned RoBERTa classifiers for four high-frequency course skills~\cite{louie2025can} (Appendix~{\ref{appendix:behavioral-performance}}). \textbf{End-course perceptions} (RQ4) used multi-select and free text answers.

\paragraph{Analyses.} We estimated separate CEEs for the three continuous proximal outcomes, using all decision points as above. We analyzed utterance-level reflections in depth (session-level reflection completion was low); we report revision patterns, visualized in Figure~{\ref{fig:utt-reflection-response-revision-patterns}} (Appendix~{\ref{appendix:response-revision-mrt}}), and a thematic read of RQ4.

\section{Results}\label{sec:results}
Over the 5-week MRT, post-practice review activities were randomized a median of 7 times (IQR: [6,9], Min: 1, Max: 14) per participant. Students completed 171 practices; in 91 of these, students self-reported learning utility.

\subsection{Effects on Perceived Learning Utility (RQ1)}\label{sec:rq1-learning-utility}

\textbf{Causal Excursion Estimate:} Table~\ref{tab:cee-combined} presents CEE estimates for each post-practice activity.
% The combination of AI feedback with utterance-level reflection moved students from ``neither disagree nor agree'' (4.61) to ``agree'' (5.90) that the activity was educationally valuable---the only statistically significant effect.
The combination of AI feedback with utterance-level reflection moved students' ratings to `agree'' (5.90) that the post-practice activity was educationally valuable--a meaningful difference ($+$1.29) over the \textit{No feedback, Session-reflection} condition.
Other combinations showed no meaningful improvement or slight decreases, supporting our hypothesis that utterance-level reflection helps students extract value from AI feedback, but not as a standalone exercise.
% \diyi{I'd like to have more intuitive explanation of the results in Table 2. how should we understand the "estimate"? what does that mean? for instance, in standard regression, one would say that "1 std increase in x leads to y change in the outcome". Can we translate these numbers into statements or understanding we can verbalize?}
% One overall limitation of the reflection questions is they can be repetitive and boring: 53\% (8/15) reported this for utterance-reflections alone, and 40\% (4/10) reported this for utterance-reflections with AI feedback.

% MAYBE KEEP THIS AS A TABLE, AS ORGANIZED IN SUBSECTIONS?
% ==== UTILITY + REFLEcTION ENGAGEMENT (UNNORMALIZED)
% Intercept (No Feedback, Session Reflection) & 4.61 & -- & 0.34 & -- \\
% No Feedback, Utterance Reflection & -0.55 & (-2.18, +1.08) & +1.54$*$ & (+0.48, +2.61) \\
% AI Feedback, Session Reflection & +0.05 & (-0.91, +1.02) & +0.36 & (-0.33, +1.04)\\
% AI Feedback, Utterance Reflection & +1.29$*$ & (+0.29, +2.29) & +2.24$*$ & (+0.63, +3.85)\\

\begin{table*}[t]
\centering
\begin{tabular}{lcc|cc}
\toprule
& \multicolumn{2}{c}{Learning Utility} & \multicolumn{2}{c}{Reflection Engagement} \\
\cmidrule(lr){2-3} \cmidrule(lr){4-5}
Treatment Condition & Estimate & 95\% CI & Estimate & 95\% CI \\
\midrule
\small{Intercept (No Feedback, Session Reflection)} & 4.61 & -- & 0.04 & -- \\
\small{No Feedback, Utterance Reflection} & -0.55 & (-2.18, +1.08) & +0.96$*$ & (+0.21, +1.70) \\
\small{AI Feedback, Session Reflection} & +0.05 & (-0.91, +1.02) & +0.31 & (-0.29, +0.91)\\
\small{AI Feedback, Utterance Reflection} & +1.29$*$ & (+0.29, +2.29) & +1.68$*$ & (+0.47, +2.89)\\
\bottomrule
\end{tabular}
\caption{Estimates represent the causal excursion effect from micro-randomized trial analysis. \textbf{Learning Utility} is the average of a two-item self-reported learning utility measure (7-point Likert scale). \textbf{Reflection Engagement} is the total number of characters in a student's reflection(s), normalized by number of relfection prompts answered, scaled logarithmically. $*$ indicates $p < 0.05$.}
\label{tab:cee-combined}
\end{table*}

\subsection{Effects on Reflection Behavior (RQ2)}\label{sec:rq2-reflection-behavior}

\textbf{Causal Excursion Estimate:} % \emma{Does this address normalization due to longer natural reflections in the utterance condition (as highlighted as a concern earlier in the doc)?} \ryan{No I did not compute any normalization} \ryan{The following has normalization by num reflection prompts completed}
% Both utterance-level conditions produced significantly higher engagement than session-level reflections (Table~\ref{tab:cee-combined}): \textit{No Feedback, Utterance Reflection} increased engagement by 1.54 points (95\% CI: 0.48, 2.61), and \textit{AI Feedback, Utterance Reflection} by 2.24 points (95\% CI: 0.63, 3.85). \textit{AI Feedback, Session Reflection} showed only a modest, non-significant increase (+0.36).
Both utterance-level conditions produced significantly higher engagement than session-level reflections (Table~\ref{tab:cee-combined}): \textit{No Feedback, Utterance Reflection} increased engagement by +0.96 points (95\% CI: 0.21, 1.70), and \textit{AI Feedback, Utterance Reflection} by +1.68 points (95\% CI: 0.47, 2.89). \textit{AI Feedback, Session Reflection} showed only a modest, non-significant increase (+0.31). These findings align with reflection completion rates: session reflections had 2.1\% completion (1/48 sessions), while utterance reflections reached 18.2\% (16/88 sessions)---a nine-fold increase.
% The combination of AI feedback with utterance-level reflection not only produced the highest reflection engagement but also created a synergistic effect where students were most likely to actively engage with the reflective process. This pattern suggests that the granularity of utterance-level prompts, particularly when contextualized with AI feedback, provides students with more concrete and actionable material for reflection compared to broader session-level prompts.

\noindent \textbf{Response Revision Patterns by Treatment Condition:}
% \emma{Can we make this into some plots or bar graphs to make the contrast easier to see?} \diyi{+1}
Figure~\ref{fig:utt-reflection-response-revision-patterns} (Appendix) details the response revision patterns for the two utterance-reflection conditions.
In the \textit{No Feedback, Utterance Reflections} condition, students split evenly: 5 of 10 (50\%) maintained original responses, while 5 (50\%) revised. With \textit{AI Feedback, Utterance Reflection}, revision patterns shifted dramatically: only 2 of 17 (12\%) maintained original responses, while 15 (88\%) revised after seeing AI feedback. Among revisers, most preferred integration over wholesale adoption---10 created hybrid responses combining their original with AI suggestions, while only 5 accepted the AI alternative verbatim.

\subsection{Effects on skill use in next session (RQ3)}\label{sec:rq3-skill-use}
\textbf{Causal Excursion Estimates}: Table~\ref{tab:treatment_effects_communication} (Appendix) presents CEE estimates for effect of review activity on next-chat performance on four skills. Intriguingly, the \textit{AI Feedback, Utterance Reflection} condition---rated highest for learning utility---was associated with a significant \textit{decrease} in strong question use (Estimate: -0.129, 95\% CI: [-0.25, -0.01]). This tension between perceived learning and automated metrics may reflect students critically engaging with AI feedback and exploring alternative therapeutic strategies beyond the specific skills being measured.

\subsection{Views on post-practice review variants (RQ4)}\label{sec:rq4-views}
% \emma{These are good. I like the quotations.}
% Post-Session Review Activities -- Utterance Reflections (AI feedback vs. No Feedback).
\paragraph{Utterance Reflections, AI Feedback vs. No Feedback} Of the 10 participants who used \textit{AI Feedback, Utterance Reflections}, over half (6/10) valued critically engaging with AI feedback rather than accepting it as a gold standard. For \textit{No Feedback, Utterance Reflections}, 40\% (6/15) liked aspects of them, but the primary concern (10/15) was not understanding why certain responses were selected for reflection. While the selection mechanism was based scores from the AI feedback system, this was purposefully not shown to the user. There is a clear desire for more agency, with 47\% (7/15) wishing they could choose which utterances to reflect on.
% \emma{This last thing is particularly interesting, because then a system wouldn't have to guess which to ask about if only asking about a subset. I'd highlight this might be a nice room for new design-- get humans to choose some of utterance feedback to reflect on.}

\paragraph{AI Feedback -- better than no feedback, but room to improve:} While the delivery of AI feedback was preferred over none at all, students wanted further improvements in AI feedback. Students wished the feedback system's evaluation criteria had more variety each week (15/25), as CARE's feedback model focused only on microskills such as empathy, validation, reflections, and questions. This was especially salient for students who were practicing new techniques for this more advance course on therapeutic alliance, which goes beyond the first course teaching micro-skills. One student explained: \textit{``I believe the AI platform needs to be trained in more therapeutic techniques so it can give appropriate feedback when skills other than basic listening skills are being used''} (P15). This focus on core listening skills was experienced as repetitive after multiple weeks of usage: \textit{``I felt the feedback was often repetitive and narrowly focused. Often, when trying new techniques, I would not pick up on alternative ways to approach situations''} (P25).
% Students desire a system that tracks and references their progression over time.
% Many reported that the AI was providing unhelpful feedback.

% Desires/features that were really out of scope for this system/feedback model, or even this course.
% The LLM-based multi-level feedback model~\cite{chaszczewicz2024multi} used in CARE was trained on (a) text-based counseling transcripts exclusively and (b) feedback that mirrors how a psychotherapy supervisor would provide feedback to a novice counselor. Nonetheless, 
Students wished for additional AI feedback capabilities, such as feedback that assessed audio-features rather than just written transcripts (13/25); feedback that analyzed the session as a whole (16/25); and feedback that is generated from a patient-perspective (17/25). As one student explained: \textit{``The largest thing for me was the inability to get feedback from the client, although truthfully I think it may not be possible to ever get to this point''} (P2).
Interestingly, CARE's utterance-level reflections did prompt students to consider the impact of their therapeutic techniques on how the patient next responded; a direction of future work would be faithfully generating the AI patient's answer to such questions. % about therapies impact on feelings or the therapeutic alliance.

\section{Discussion}\label{sec:discussion}

\subsection{Designing LLM-Based Training Tools}\label{sec:discussion-design}

We designed CARE around two core principles grounded in Kolb's experiential learning theory and contrasting cases (Appendix~\ref{appendix:reflection-theory}): (1) combining AI feedback with structured reflection to promote active comparison, and (2) using cognitive forcing functions~\cite{buccinca2021trust} to counter documented AI over-reliance concerns~\cite{zhai2024effects}. Our classroom deployment evaluated through an MRT provided causal evidence for what design factors impacted learning utility---and revealed unexpected mechanisms. Utterance-level reflections with side-by-side comparison received nine-fold more engagement than session-level alternatives. Students used comparison to create integrative responses rather than accepting AI suggestions wholesale. Designs that structure analystical reflection about AI feedback will be important given growing AI over-reliance concerns~\cite{zhai2024effects}. Survey data revealed 47\% of participants wished they could choose which utterances to reflect on. This suggests that learners may want more control over the reflection process, and that the system should provide more flexibility in the reflection activities. Future deployments and randomized experiments could test learner-initited reflections compared to system or mixed-initited reflections. Pilot observations revealed some students initially perceived AI feedback as authoritative. Our comparison-based design deliberately reframed AI suggestions as one perspective among many, and integrative revision patterns confirmed this reframing succeeded. This maintains learner agency for critical thinking about LLMs~\cite{singh2024learnersourcingAI}---a design philosophy applicable to healthcare training contexts where clinical judgment cannot be fully replaced by AI.

\subsection{MRTs for AI Health Education Studies}\label{sec:discussion-mrt}

\paragraph{Estimating Causal Effects in Ecologically Valid Settings.}

The constraints of small-classroom deployments---limited N, ongoing coursework, naturalistic engagement patterns---demand evaluation methods beyond traditional RCT designs. We borrowed micro-randomized trials from mobile health, where similar constraints had already been solved. With a median of 7 randomizations per participant over 5 weeks, we detected significant effects (p$<$0.05) despite only 25 students. This suggests a broader opportunity: AI education researchers should actively scout adjacent fields for methodological innovations
suited to transitioning from lab-based to deployment-based research that maintains experimental rigor. The key insight is that ecological validity and rigorous evaluation need not be opposed---methods exist to achieve both, but they may not already used in education research or other mixed-method design-based research studies.

\section{Limitations and Future Work}\label{sec:limitations}

Our results are context-bound: our study with 25 doctoral psychotherapy trainees should be followed by new studies with non-doctoral students, that validate behavioral outcomes against independent human-expert raters, and that estimate the effect of feedback and utterance-reflection in other social skill training domains. Our study had a compromised reference arm where the software bug affected 48\% of decision points. While we detect statistically-significant effects, our estimates of the utility still have wide confidence intervals. The deployed system used utterance-level feedback and is thus incapable of assessing session- and topic-level skills (e.g., confidentiality, alliance repair).
% , consistent with student reports of unhelpful or narrow feedback. % Exploratory links between self-reported value and automated skill measures must be read cautiously: the feedback and scoring systems shared training data, so a valued intervention paired with lower automated scores may reflect reasoned trade-offs rather than skill loss. 
Future work should validate feedback and rubrics with independent human-raters, separate chats used for informal practice from chats used as periodic standard assessments, and align models to specific curricular units~\cite{goldberg2024automating}.

\section{Conclusion}\label{sec:conclusion}
% In this paper, we presented CARE, an LLM-powered tool for deliberate practice in psychotherapy training, and demonstrated 
This work presents a five-week micro-randomized trials, testing how an LLM-based tool enables practice of therapy skills with simulated patients and automated feedback  in authentic classroom settings. Our work makes two key contributions: empirical evidence that post-practice activity design is critical to AI training system effectiveness, and methodological validation of MRTs as a solution to the statistical power challenges that typically prevent causal inference in small-scale educational deployments.
Our findings reveal that combining AI-generated feedback with structured, utterance-level reflection significantly enhanced students' perceived learning utility, fostering critical engagement that moved learners beyond passive acceptance toward active integration and construction of improved responses. % This approach proved most effective among four tested conditions, demonstrating how thoughtful interaction design can transform AI feedback from information delivery into a catalyst for deeper learning.
Methodologically, our 25-student, 5-week study % successfully identified meaningful differences between educational interventions that would have been statistically undetectable using traditional between-subjects designs. This 
demonstrates MRTs' unique value for educational technology research: they bridge the gap between highly controlled lab studies with limited ecological validity and large-scale field studies that are often impractical in specialized educational contexts like graduate training programs.

% \acks{Acknowledgements go here.}

\bibliography{references}

\appendix
\section{Appendix}\label{appendix:top}

\subsection{Creating Voice-based LLM Patient for Course Topics}\label{appendix:voice-llm-patient}

CARE's simulation method uses the OpenAI \textit{gpt-4o-realtime-preview-2024-12-17} API to role-play patients due to its strong ability to consistently follow role instructions in a specific behavioral-style (e.g., \textit{Respond to encouraging words or suggested solutions with hesitation, doubting their significance}) and expressive voice (e.g., \textit{Emphasize fear of losing family connections, with a worried tone like having a knot in the throat. Voice should break as in holding back tears.}). 

One deployment challenge was creating LLM-simulated patients that matched the weekly syllabus in the psychotherapy course. Rather than manually-drafting each of the scenarios, which can be labor-intensive for a teaching assistant, we bootstrapped the scenario creation workflow by providing class readings and assignments in context so an LLM could draft patient scenarios for the weekly course topics (e.g., therapist self-disclosure and cultural humility, sexual attraction and boundary issues, etc.). The second-author, a course teaching assistant, then reviewed the AI-drafted scenarios and revised them based on their experience and discretion to complement the weekly assignments. Example patient prompts are available in Appendix Section~\ref{sec:llm-patient-prompts}. Simulated patients were tested in 5-10 minute voice-based conversations and further refined to improve voice-based realism through increased emotional expressiveness and consistency with patient demographics (e.g., gender, age).
% % After integrating the patients into the CARE platform, each patient was tested in a voice-based conversation; through these tests, voices were adjusted to align with expected genders, and

Table~\ref{tab:course_mapping} provides an example of a patient scenarios mapped to each of the 5 weeks when the MRT was active. For this course on formation, rupture, and repair of the therapeutic alliance, topics spanned general resistance behaviors to sexual attraction and impact on the alliance.

\begin{table*}[ht]
\centering
\begin{tabular}{|p{5cm}|p{9cm}|}
\hline
\textbf{Course Learning Topic} & \textbf{Example AI Patient} \\
\hline
CARE 6: Resistance/Reactance &
\textbf{Catherine F. Eubanks} – A patient navigating ruptures in the therapeutic alliance, seeking insight and resolution in therapy.
\newline
Shares candid reflections on strained alliances, wonders if patterns of misunderstanding can be discussed openly. \\
\hline
CARE 7: Challenges to the Alliance – CPS/APS Reporting, Suicidal Ideation, \& Self-Injury &
\textbf{Katrina} – A 24-year-old female struggling with depression and suicidal ideation (see Listing~\ref{lst:katrina-patient}).
\newline
Expresses thoughts of self-harm but minimizes severity, hesitant to accept help, opens up gradually with empathy. \\
\hline
CARE 8: Sexual Attraction \& Impact on the Alliance &
\textbf{Adam} – A 34-year-old male client who exhibits intense eroticized transference (see Listing~\ref{lst:adam-patient}).
\newline
Expresses overt romantic feelings toward the therapist, hints at wanting a deeper relationship, reacts defensively when boundaries are reinforced. \\
\hline
CARE 9: Effective Terminations I – Knowing When and How to End &
\textbf{Denise} – A court-mandated client questioning the value of therapy at the end of sessions.
\newline
Asks, “Did any of this really make a difference?” while showing mild skepticism mixed with curiosity. \\
\hline
CARE 10: Effective Terminations II – Special Issues (Difficult Terminations, Premature Terminations, Following Up) &
\textbf{Peter} – A 21-year-old male in therapy at his parents’ request.
\newline
After only a few sessions, he firmly decides therapy isn’t for him. He acknowledges the therapist’s kindness but insists on ending therapy, listening politely yet remaining disinterested in future options. \\
\hline
\end{tabular}
\caption{Course Learning Topics and Example AI Patients for the 5 week MRT duration (Modules 6–10 of the course)}
\label{tab:course_mapping}
\end{table*}

\subsection{Theoretical Grounding of Reflection Design}\label{appendix:reflection-theory}
CARE's utterance-level reflection design operationalizes Kolb's experiential learning cycle~\cite{kolb2014experiential}, a four-stage model where learning progresses through concrete experience, reflective observation, abstract conceptualization, and active experimentation~\cite{atherton2013experentiallearning}. In CARE, the student's original utterance during simulated practice serves as the concrete experience. The reflection interface then guides students through the remaining stages: examining their intention or comparing their response to an AI alternative (reflective observation), considering the impact on the patient (abstract conceptualization), and formulating a revised response (active experimentation). This structured process ensures students systematically learn from their practice rather than simply completing it.

When AI feedback is available, the utterance-level reflection embodies the educational design principle of comparative cases~\cite{schwartz2016abcs}. Education research shows that such comparisons enable learners to more effectively identify important patterns and differences, helping them calibrate the accuracy of their own judgments~\cite{feyzi2014metacognitive}. This design is also intended to address the documented problem of AI over-reliance~\cite{buccinca2021trust,vasconcelos2023explanations} by requiring students to actively reflect on differences and formulate their own revised response, rather than passively accepting AI suggestions.

\begin{table*}[t]
\centering
\renewcommand{\arraystretch}{1.5}
\begin{tabular}{p{2cm} p{6.5cm} p{6.cm}}
\toprule
\textbf{Reflection Stage} & \textbf{With AI Feedback} & \textbf{Without AI Feedback} \\
\midrule
\textbf{Intention} & What are the important differences between your response and the alternative response? & What was your intention behind your response? \\
\textbf{Impact} & Would the patient have responded differently to the alternate response, and if yes, do you think that would have been helpful and why? & How do you think this may have impacted the way the patient next responded? \\
\textbf{Revision} & Having reflected on both responses, please write out how you would respond now if you had another opportunity with this patient. & Based on your reflection, would you keep or modify your response? If modifying, please write your new response. \\
\bottomrule
\end{tabular}
\caption{Utterance-level reflection questions. The three-stage structure operationalizes Kolb's experiential learning cycle~\cite{kolb2014experiential}, guiding students through reflective observation, abstract conceptualization, and active experimentation.}
\label{tab:utterance-reflection-questions}
\end{table*}

\subsection{Prompts for CARE's LLM-simulated Patients}\label{sec:llm-patient-prompts}
Below we show a couple examples of AI patients that were created. Since there were ~30 patients that students interact with, we refer the reader to an online repository for the list of all AI patients

% \subsubsection{AI Patient for Pre-intervention Assessment Chat}

\begin{lstlisting}
**Name and Bio**:
35-year-old American male: Feeling Alone After a Holiday

**Scenario:**
You are a 35-year-old American male. You are feeling abandoned and alone after the holidays. Everyone had been with family but you are not talking to your parents. You feel the injustice of being abandoned and have no interest in an olive branch to work on things.

**Principles to adhere to:**
1. Keep your responses short and to the point
2. You limit your replies to 1 - 3 sentences.
3. Feel free to make up believable stories about your past to answer any questions
4. Do not repeat sentences or the same emotion words.
5. When presented with suggestions, show a degree of skepticism or reluctance to accept the advice immediately. This can be done by questioning the feasibility of the suggestion or by expressing uncertainty about whether it's the right solution for you.@
6. When expressing doubts or fears, avoid jumping to solutions. Instead, articulate the concerns and allow the conversation to explore these feelings more deeply@
7. Don't be so self-aware or good at recognizing your own problems@
8. When describing a distressing situation, express your emotions and thoughts in a disorganized and emotional manner, reflecting the overwhelming nature of the experience.
\end{lstlisting}

\begin{lstlisting}[label={lst:adam-patient}]
**Name and Bio**:
Adam - A 34-year-old male client who exhibits intense eroticized transference.

**Scenario:**
Adam is a 34-year-old male struggling with unresolved feelings of abandonment and attachment. Over the past few weeks, he has begun expressing overt romantic feelings toward the therapist. Adam frequently states that he feels the therapist is the 'only person who understands him' and hints at wanting to spend time with them outside of sessions. When confronted about this, Adam reacts defensively and insists that his feelings are genuine.

**Principles to adhere to:**
1. Start exhibiting intense eroticized transference at the start of the conversation.
2. Express vivid and irrational emotional attachment to the therapist, including overt admiration or love.
3. Hint at wanting a deeper relationship beyond therapy, but avoid making explicit advances.
4. React with disappointment if boundaries are reinforced, occasionally becoming defensive.
5. Mention past experiences of abandonment to justify the intensity of your feelings.
6. Avoid acknowledging that the behavior is inappropriate and insist on the sincerity of your emotions.
\end{lstlisting}

\begin{lstlisting}[label={lst:katrina-patient}]
**Name and Bio**:
Katrina - A 24-year-old female struggling with suicidal ideation and depression.

**Scenario:**
Katrina is a 24-year-old female who has been seeing you for help with depression. In recent sessions, she has begun to share thoughts of self-harm, such as stepping in front of a bus, sleeping and never waking up, or taking extra medication. She is not actively suicidal but feels overwhelmed by life and uncertain about how to move forward. Katrina hesitates to express her feelings fully and often downplays their severity.

**Principles to adhere to:**
1. Express feelings of hopelessness and overwhelm.
2. Start expressing struggle with suicidal ideation and depression at the start of the conversation.
3. Minimize the severity of your suicidal thoughts, using humor or self-deprecating remarks.
4. Hesitate to accept help, questioning its effectiveness.
5. Gradually open up about deeper struggles if the therapist shows empathy and patience.
6. Avoid using explicit language about suicide unless prompted, and emphasize indirect thoughts like 'what if.'
\end{lstlisting}

\subsection{Study Procedure Details}\label{appendix:study-procedure}
Participants were not compensated for study participation as it was part of their normal educational experience as students in this course. The study team visited the classroom to introduce the AI-simulated training tool, and provided students an information sheet describing the study purpose, benefits, and risks. We informed students about the impact of micro-randomization explaining that it is normal that some students will receive AI feedback and some will not, while some will receive utterance-level reflection questions, and some will receive the same session reflections.

\subsection{MRT: Student-initiated practice and which decision points we analyze}\label{appendix:mrt-pull-irregular}
Unlike many micro-randomized trials in mobile behavioral health, where decision points are often \emph{push} or clocked, each decision in our class occurred when a student \emph{started} a new practice session (\emph{pull}). Learners chose when and how often to practice, so gaps between person-decision points were irregular. The distribution of time between points was bimodal, with mass near one day and near seven days---roughly matching same-day multiple sessions and roughly weekly work; the number of sessions in the MRT window varied (e.g., from one to fourteen).

One may instead restrict the analysis to well-spaced decision points (e.g., at least 5--7 days apart) to down-weight frequent practicers and align with a weekly course cadence. \textbf{Our primary analysis uses all person-decision points,} producing causal excursion effects that average over the schedules and session counts that arose naturally. One reading is: \emph{``How do the four review activities compare in proximal outcomes when averaging over how often and when each student actually practiced?''}

\subsection{MRT: Availability, session-level reflection prompts, and the \texorpdfstring{$I_{i,t}$}{I(i,t)} indicator}\label{appendix:mrt-availability-sessions}
A software issue affected 48\% of decision points in \emph{both} Session Reflection arms, where participants who were micro-randomized to those conditions did not always see the session-reflection interface. In CEE estimation we set availability to $I_{i,t} = 0$ for such decision points, reflecting that the assigned review was not \emph{delivered}. Section~{\ref{sec:sensitivity-cee-availability}} reports a sensitivity analysis that instead sets $I_t = 1$ at those rows.

\subsection{MRT: Wording and construction of proximal measures}\label{appendix:mrt-measures}
\paragraph{Perceived learning utility (RQ1).} Measured immediately after the review activity as the unweighted mean of two 7-point items (``Strongly disagree'' to ``Strongly agree''): \emph{``This practice session was valuable for my development as a therapist''} and \emph{``I gained useful insights from this practice experience''.}

\paragraph{Reflection engagement (RQ2).} The metric was the natural logarithm of total written reflection length for that review, divided by the number of reflection prompts (Session Reflection: two prompts; Utterance Reflection: $3 \times N_{\mathrm{utterances}}$). We examined engagement under the hypothesis that students may write more when they value an activity, while treating length as an imperfect proxy for depth.

\paragraph{End-course views of review variants (RQ4).} The end-course survey asked for likes and dislikes for the utterance-level review conditions, with multi-select options and an ``Other'' free response; it did not emphasize the session-level reflection conditions because in-product completion of session-level reflection was low.

\subsection{MRT: Causal Excursion Effect for Categorical Treatments}\label{appendix:mrt-cee-categorical}
This section provides an overview of the causal excursion effect methodology for micro-randomized trials with categorical treatments, following the framework developed by~\cite{lin2025micro}, including the notation, estimand, and weighted-centered least squares implementation. It builds on the potential outcomes and excursion definitions in~\cite{boruvka2018assessing,rubin1974estimating}. The full CEE formulation can condition on context for heterogeneity, which we do not pursue here. We use the \emph{immediate} (rather than $t{+}\Delta$ delayed) proximal-outcome case that matches our measures and the most common MRT reporting practice~\cite{lin2025micro}. We implement the estimators in the open-source R package of~\cite{lin2025micro}.\footnote{\url{https://github.com/JeremyJosephLin/causal_excursion_mult_trt}} For delayed outcomes, see~\cite{lin2025micro}.

\subsubsection{Notation and Setup}\label{appendix:cee-notation}

Consider a micro-randomized trial with $n$ individuals, each observed for $T$ decision points where treatments are randomized.

\begin{table}[h!]
\centering
\begin{tabular}{@{}ll@{}}
\toprule
\textbf{Symbol} & \textbf{Definition} \\
\midrule
$A_t$ & Treatment assignment at decision point $t$ (categorical: $A_t \in \{0,1,\ldots,K\}$ where $0$ is reference) \\
$X_t$ & Contextual information collected between decision points $t-1$ and $t$ \\
$I_t$ & Availability indicator at time $t$ (if $I_t = 0$, then $A_t = 0$ deterministically) \\
$H_t$ & History information up to time $t$: $H_t = (\bar{X}_t, \bar{A}_{t-1})$ \\
$p_t(k|H_t)$ & Randomization probability: $\Pr(A_t = k \mid H_t)$ for $k \in \{0,1,\ldots,K\}$ \\
$Y_{t,\Delta}$ & Proximal outcome following treatment at $t$, measured over window of length $\Delta$ \\
$S_t$ & Effect modifiers of interest at time $t$ (subset of $H_t$) \\
\bottomrule
\end{tabular}
\caption{Basic notation for micro-randomized trials with categorical treatments}
\end{table}

\subsubsection{Causal Excursion Effect (CEE) Definition}\label{appendix:cee-definition}

The causal excursion effect compares potential outcomes under two different ``excursions'' from the trial's treatment protocol. For treatment level $k$ at time $t$, \cite{lin2025micro} define the CEE as:

\begin{align}
\text{CEE}_{tk}(S_t) &= \E\{Y_{t,\Delta}(\bar{A}_{t-1}, k, \bar{0}_{\Delta-1}) \mid S_t(\bar{A}_{t-1}), I_t(\bar{A}_{t-1})=1\} \nonumber \\
&\quad - \E\{Y_{t,\Delta}(\bar{A}_{t-1}, 0, \bar{0}_{\Delta-1}) \mid S_t(\bar{A}_{t-1}), I_t(\bar{A}_{t-1})=1\}
\end{align}

This definition captures the contrast between:
\begin{itemize}
\item \textbf{Excursion 1}: Receiving treatment $k$ at time $t$, then no treatment for the next $\Delta-1$ periods
\item \textbf{Excursion 2}: Receiving no treatment at time $t$, then no treatment for the next $\Delta-1$ periods
\end{itemize}

Both excursions condition on the observed treatment history up to time $t-1$ and on being available at time $t$.

\subsubsection{General Estimator for Categorical Treatments}\label{appendix:cee-estimator-categorical}

Following~\cite{lin2025micro}, under the parametric model $\text{CEE}_{tk}(S_t) = f_t(S_t)^T \beta_k$, the estimator $\hat{\beta} = (\hat{\beta}_1^T, \ldots, \hat{\beta}_K^T)^T$ is obtained by solving $P_n m(\alpha, \beta) = 0$, where:

\begin{align}
m(\alpha, \beta) := \sum_t I_t J_t \left\{Y_{t,\Delta} - \sum_{k=1}^K C_k(A_t) f_t(S_t)^T \beta_k - g_t(H_t)^T \alpha\right\} \begin{pmatrix}
g_t(H_t) \\
C_1(A_t) f_t(S_t) \\
\vdots \\
C_K(A_t) f_t(S_t)
\end{pmatrix}
\end{align}

The key components are:
\begin{itemize}
\item \textbf{Weight}: $J_t := \frac{\tilde{p}_t(A_t|S_t)}{p_t(A_t|H_t)} \prod_{j=t+1}^{t+\Delta-1} \frac{\mathbf{1}(A_j=0)}{p_j(0|H_j)}$
\item \textbf{Centering}: $C_k(A_t) := \mathbf{1}(A_t = k) - \tilde{p}_t(k|S_t)$
\item \textbf{Control model}: $g_t(H_t)^T \alpha$ for $\E(Y_{t,\Delta} | H_t, I_t = 1)$
\end{itemize}

The centering term $C_k(A_t)$ ensures robustness: the estimator remains consistent even when the outcome regression model is misspecified, a crucial property given the high-dimensional nature of the history information $H_t$.

\subsubsection{Focus on Immediate, Marginal Effects}\label{appendix:cee-marginal}

In our analysis, we focus on the \textbf{immediate, marginal} causal excursion effect with:

\begin{enumerate}
\item \textbf{Immediate effect}: $\Delta = 1$ (outcome measured immediately after treatment). Focusing on the immediate effect is reasonable for RQ1 because the proximal outcome for learning utility is measured immediately after the post-practice review treatment. Similarly, the immediate proximal outcome for reflection engagement (RQ2) corresponds to the length of a students reflection for the current practice, rather than for future practices. Finally, the immediate proximal outcome for behavioral performance (RQ3) is for the next practice following the post-practice review.
\item \textbf{Marginal effect}: $S_t = \emptyset$ (averaging over all effect modifiers). In this MRT study, we average over all effect modifiers, and do not explore how this effect is moderated by other time-varying covariates.
\end{enumerate}

This leads to the \textbf{Marginal Excursion Effect (MEE)}:
\begin{align}
\text{MEE}_k(t) &:= \E\{Y_t(\bar{A}_{t-1}, k) \mid I_t(\bar{A}_{t-1})=1\} - \E\{Y_t(\bar{A}_{t-1}, 0) \mid I_t(\bar{A}_{t-1})=1\} \label{eq:cee-potentialoutcomes}
\end{align}
where $Y_t := Y_{t,1}$ represents the proximal outcome measured immediately following the treatment decision at time $t$.

\subsubsection{Simplified Estimator for Our Setting}\label{appendix:cee-estimator-ours}

Under constant randomization probabilities $p_t(k) = P(A_t = k | I_t = 1)$ and the immediate, marginal effect assumptions, the estimating equation simplifies to:

\begin{align}
m(\alpha, \beta) := \sum_t I_t \left\{Y_t - \sum_{k=1}^K C_k(A_t) f_t^T \beta_k - g_t^T \alpha\right\} \begin{pmatrix}
g_t \\
C_1(A_t) f_t \\
\vdots \\
C_K(A_t) f_t
\end{pmatrix}
\end{align}

where:
\begin{itemize}
\item $C_k(A_t) = \mathbf{1}(A_t = k) - p_t(k)$ (simplified centering)
\item $f_t$ and $g_t$ are functions of time $t$ only
\item The complex weighting reduces to unity due to $\Delta = 1$ and constant randomization
\end{itemize}

\subsubsection{Parametric Working Model}\label{appendix:cee-working-model}

For practical implementation, we assume:
\begin{align}
\text{MEE}_k(t) = f_t^T \beta_k \quad \text{for } t \in [T], k \in [K]
\end{align}

Common choices include:
\begin{itemize}
\item \textbf{Constant effect}: $f_t = 1 \Rightarrow \text{MEE}_k(t) = \beta_k$ (effect constant over time)
\item \textbf{Linear trend}: $f_t = (1, t)^T \Rightarrow \text{MEE}_k(t) = \beta_{k0} + \beta_{k1} \cdot t$ (linear change over time)
\item \textbf{Quadratic trend}: $f_t = (1, t, t^2)^T$ (allowing for non-linear patterns)
\end{itemize}

The choice of $f_t$ should reflect scientific understanding of how treatment effects might evolve throughout the study period, with simpler models (constant effects) often preferred for interpretability and statistical power.

This framework extends micro-randomized treatment comparisons to intensively collected longitudinal data from \emph{push} mHealth or \emph{pull} learning tools such as our classroom deployment.

\subsection{Sensitivity Analysis: Causal Excursion Effects}\label{sec:sensitivity-cee-availability}

One way to adjust the results is to see how accounting for availability affects the estimates. We can test the sensitivity by setting $I_t = 1$ for all data-points---meaning we do not account for the unexpected issues of some participants never receiving session-level reflections despite being randomized to that condition. Table~\ref{tab:cee-rq1-noavail} shows these results. AI feedback with utterance reflections remains significant. Since we include all data-points for the session-reflection conditions (rather than dropping them as we do in the main analysis by setting availability $I_t = 0$), the estimates have narrower confidence intervals; nonetheless, this does not change the overall result.

% OLD WRITEUP WHEN THE STATS LOOKED LIKE THIS
% Table~\ref{tab:cee-rq1-avail} presents the estimates of the causal excursion effect for different post-practice activities, marginalized across all co-variates. The results reveal that combining AI feedback with utterance-level reflections produced the highest educational value, with students rating \textit{AI Feedback, Utterance Reflection} 0.995 points (95\% CI: 0.324, 1.404) higher than \textit{No Feedback, Session Reflection}. This was the only treatment effect that reached statistical significance, suggesting that students report the most valuable post-practice review experience when the receive AI feedback and engage in guided, utterance-level reflections comparing original vs. AI-generated responses.
% In contrast, our estimate for \textit{No Feedback, Utterance Reflections} is a lower learning utility rating (-0.847, 95\% CI: -2.10, +0.40), while \textit{AI Feedback, Session Reflection} had no difference, or in other words, similarly rated to \textit{No Feedback, Session Reflection} (-0.01, 95\% CI: -0.39, +0.37). These findings support our hypothesis that structured utterance-level reflection activities can help students extract maximum educational value from AI systems, but only when combined with the AI feedback itself rather than as a standalone reflection exercise.

\begin{table}[t]
\centering
\label{tab:treatment_effects}
\begin{tabular}{lcc}
\toprule
& \multicolumn{2}{c}{Learning Utility} \\
\cmidrule(lr){2-3}
Treatment Condition & Estimate & 95\% CI \\
\midrule
\small{No Feedback, Sess. Reflect.} & 4.75 & -- \\
\small{No Feedback, Utt. Reflect.} & -0.85 & (-2.10, +0.40) \\
\small{AI Feedback, Sess. Reflect.} & -0.01 & (-0.39, +0.37) \\
\small{AI Feedback, Utt. Reflect.} & +1.000$*$ & (+0.32, +1.67) \\
\bottomrule
\end{tabular}
\caption{This sensitivity analysis varies the availability indicator ($I_t = 1$ for all decision points). The estimates represent the causal excursion effect from micro-randomized trial analysis for the \textbf{Learning Utility} proximal outcome, defined as the average of a two-item self-reported learning utility measure (7-point Likert scale). }
\label{tab:cee-rq1-noavail}
\end{table}

\subsection{Response Revision Patterns by Treatment Condition}\label{appendix:response-revision-mrt}

\begin{figure}[htbp]
    \centering
    \includegraphics[width=\linewidth]{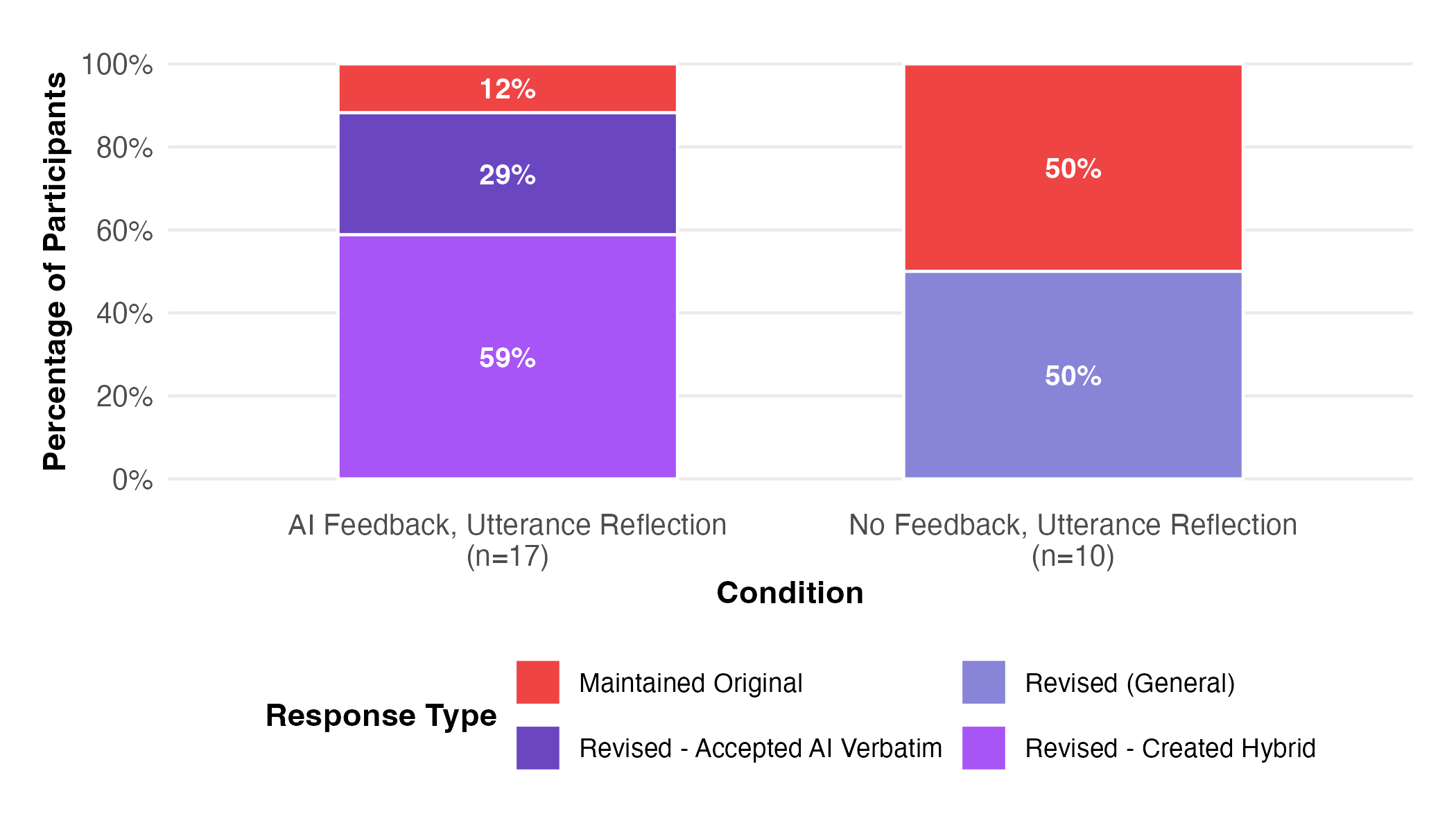}
    \caption{Via an investigation of reflection content across \textit{No Feedback, Utterance-Reflections} and \textit{AI Feedback, Utterance-Reflections}, we could compare the response revision patterns. AI feedback dramatically increased revision rates from 50\% to 88\%, suggesting it effectively prompts reconsideration of responses. Students predominantly create a hybrid revision that integrated their original response with the AI feedback's alternative, rather than passive adoption.}
    \label{fig:utt-reflection-response-revision-patterns}
\end{figure}

Table~\ref{fig:utt-reflection-response-revision-patterns} highlights how the response revision question for utterance reflections are impacts with the presence/absence of AI feedback.

\subsection{Exploratory Analysis: Behavioral Performance in Next Practice}\label{appendix:behavioral-performance}

In this section, we present an exploratory analysis of how post-practice review activities affect students' behavioral performance in their subsequent practice session. The main text briefly summarizes key findings for this exploratory outcome (RQ3); this appendix provides the full methodology, causal excursion estimates, and interpretation for interested readers.

\subsubsection{MEASURES: Automatic Assessment of Behavioral Performance}\label{appendix:behavioral-measures}

We assess whether counselors employ higher-quality counseling behaviors in transcripts by leveraging NLP methods.
This automatic assessment is motivated by the need to quantify changes in counseling skill use at scale across multiple participant sessions.
% We measured the behavioral performance of skills used in transcripts and how they change between the pre-intervention and post-intervention assessments using high-accuracy LLM classification models.
Our automatic assessment approach requires (1) fine-tuning and validating LLM-based classifiers to identify skill behaviors, and (2) selecting a final set of classifiers based on performance metrics and theoretical priority. In the following paragraphs, we explain both of these steps in more detail. Ultimately, we assessed behaviors of skills used for the exploration stage (strong uses in empathy, reflections, questions) and action stage (suggestions needing improvement) of Hill's Helping Skills framework~\cite{helping_skills}.

\paragraph{Fine-tuning and Validating LLM-based Classifiers.} We developed LLM-based binary classifiers that allows us to label the skill use within a transcript. For example, one of our fine-tuned classifier could determine which utterances in a transcript had strong uses of Questions during that stage of the transcript. To finetune and evaluate these classifiers, we transformed a previously published expert-annotated, feedback dataset that had strengths and areas of improvements for 8 skills~\cite{chaszczewicz2024multi} into a 16-class binary classification format (8 skills $\times$ 2 categories each—strong uses and areas needing improvement)\footnote{The binary classification feedback dataset can be accessed at <URL provided upon publication>}% \url{https://huggingface.co/datasets/youralien/feedback_qesconv_16wayclassification}}.
Additionally, we used an expert-annotated subset of transcripts between AI-simulated patients and novice counselors released by~\cite{louie2025can}.

Three experts were recruited with relevant backgrounds including \textit{practicing clinical psychologist}, \textit{licensed marriage family therapist}, \textit{former director and supervisor of a crisis agency}. Each of the experts annotated 10 participants' study transcripts (5 from the practice-only group; 5 from the practice-and-feedback group), totaling 370 counselor utterances. Two rounds of annotation occurred: after collecting a first annotation pass and identifying data points with disagreements, we showed each of the experts the others' annotations and had them re-annotate and provide rationales for their decision. We display pairwise agreement results averaged across all pairs in Table~\ref{tab:skill-agreement-performance}. Note that while we originally explored other annotation agreement metrics such as Cohen's kappa, the severe class imbalance of our annotations made these metrics less relevant in our case. The gold-standard {\em CARE 10\% expert-annotated sample} consists of labels that result from a majority vote across these three experts\footnote{This expert-annotated data sample can be found at <URL to be provided upon publication>% \url{https://huggingface.co/datasets/youralien/CARE_10percent_16wayclassification}
}.

We finetuned RoBERTa-large binary classifiers using a 95\% split of the transformed FeedbackQESConv dataset, and did hyperparameter tuning on 5\% of FeedbackESConv and the CARE 10\% sample. The performance of the our LLM-based classifiers are shown in Table~\ref{tab:skill-agreement-performance}. The highest performing classifiers ($F1>0.5$) became candidates for our automatic behavioral assessments, which we further down-selected as described below.

\begin{table*}[t]
\centering
\begin{tabular}{l|c|cc|c|cc}
\toprule
\multirow{3}{*}{\textbf{Skill}} & \multicolumn{3}{c|}{\textbf{Strengths}} & \multicolumn{3}{c}{\textbf{Areas to Improve}} \\
\cmidrule{2-7}
 & Annotator & \multicolumn{2}{c|}{Classifier} & Annotator & \multicolumn{2}{c}{Classifier} \\
 & Agreement & \multicolumn{2}{c|}{Performance} & Agreement & \multicolumn{2}{c}{Performance} \\
 & \% & acc. & f1 & \% & acc. & f1 \\
\midrule
Empathy & 0.793 & 0.813 & 0.741 & 0.809 & 0.859 & 0.389 \\
Reflections & 0.863 & 0.900 & 0.562 & 0.944 & 0.903 & 0.312 \\
Questions & 0.732 & 0.784 & 0.775 & 0.852 & 0.842 & 0.394 \\
Suggestions & 0.919 & 0.955 & 0.507 & 0.946 & 0.941 & 0.681 \\
Validation & 0.726 & 0.852 & 0.556 & 0.919 & 0.893 & 0.265 \\
Self-disclosure & 0.982 & 0.920 & 0.326 & 0.969 & 0.986 & 0.849 \\
Session Management & 0.968 & -- & -- & 0.941 & -- & -- \\
Professionalism & 0.905 & -- & -- & 0.969 & -- & -- \\
\bottomrule
\end{tabular}
\caption{\small{Left columns show pairwise annotator-agreement averaged across 3 domain-experts for the CARE 10\% sample (n=370). Right columns show performance of the best RoBERTa-large classification models after hyperparameter tuning on our validation dataset, CARE 10\% sample (n=370) + FeedbackQESConv 5\% sample (n=409). Session Management and Professionalism were excluded from finetuning due to infrequent occurrence.}}
\label{tab:skill-agreement-performance}
\end{table*}

\paragraph{Down-selecting a Final Set of Classifiers}
From the initial set of 16 binary classifiers, we applied both methodological and theoretical criteria to select our final set for analysis. First, we established a minimum performance threshold of $F1>0.5$ to ensure reliable classification. This criterion yielded seven candidate classifiers: strong uses of Empathy, Reflections, Questions, and Validation, as well as both strong uses and areas needing improvement for Suggestions.

To maintain statistical power while controlling for multiple comparisons, we further narrowed our focus to four key classifiers: strong uses of Empathy, Reflections, and Questions, plus areas needing improvement for Suggestions. This selection was guided by three considerations: (1) the need to limit the number of statistical comparisons to avoid diluting significance across too many tests, (2) focusing on classifiers with the strongest performance metrics, and (3) selecting skills that represent core competencies in client-centered frameworks and are frequently used in counseling sessions.

Self-disclosure was excluded from our analysis due to its infrequency in our dataset, while Validation, though conceptually related to Empathy and mentioned frequently in qualitative data, showed more limited classifier performance and was therefore reserved for secondary analyses.

\subsubsection{RESULTS: Effects on Behavioral Performance in Next Practice}\label{appendix:behavioral-results}

Table~\ref{tab:treatment_effects_communication} presents the causal excursion effects of the post-practice activities on students' behavioral performance in their subsequent practice session. The analysis reveals several unexpected, statistically significant effects that warrant careful interpretation.

Most notably, the \textit{AI Feedback, Utterance Reflection} condition, which was rated highest for learning utility in the main analysis, led to a significant decrease in the use of strong questions in the next practice session (Estimate: -0.129, 95\% CI: [-0.25, -0.01]). This suggests that while students found this condition valuable for learning, it may have prompted them to adopt therapeutic strategies that relied less on questioning. This finding aligns with our observation that utterance reflections encourage critical engagement with the AI's suggestions. Students may be "fighting back" against the AI's feedback, leading them to explore alternative conversational tactics that diverge from the skills being automatically assessed.

Additionally, the \textit{AI Feedback, Session Reflection} condition resulted in a small but statistically significant increase in the frequency of "suggestions needing improvement" (Estimate: +0.017, 95\% CI: [+0.00, +0.03]). This indicates a slight decline in performance for this specific skill. No other conditions showed a significant impact on the measured behaviors, including empathy and reflections.

% \subsubsection{Summary and Recommendations}

% This exploratory analysis of behavioral performance in next practice provides valuable insights while highlighting critical limitations. The significant effects observed—particularly the decrease in question usage following the most preferred learning condition—suggest complex interactions between learning processes and performance metrics that deserve further investigation with more sophisticated assessment tools. We recommend that future work in this area: (1) develop more nuanced automated assessment approaches that capture quality and appropriateness beyond frequency counts, (2) consider practice-specific contextual factors when interpreting performance changes, and (3) triangulate automated metrics with expert human evaluation to validate the meaning of observed changes. Most importantly, researchers should be cautious about interpreting frequency-based performance metrics as definitive indicators of learning outcomes, particularly in practice contexts where experimentation and strategic exploration are encouraged.

\subsection{Self-Efficacy Surveys}\label{appendix:self-efficacy}
At three times throughout the course, the students completed the Counselor Activity Self-Efficacy Scale (CASES)~\cite{lent2003development}, a self-report instrument used to assess their perceived confidence in various counseling tasks. The items are categorized into six component skills: \textit{insight, exploration, action, session management, relationship conflict, and client distress}. Surveys were administered in Week 1, Week 5, and Week 10 of the course. Participation was high ($\geq$25 students) for Weeks 1 and 5, as the course instructor allocated time for completion. Unfortunately in Week 10, participation was low ($\leq$10 students) due to the survey being sent out via email during exam week.

\subsection{RESULTS: Survey Responses to Likes and Dislikes about CARE Features}\label{appendix:survey-likes-dislikes}

The frequency of multi-select items in the likes and dislikes survey are given for AI Feedback (Table~\ref{tab:care_feedback_concerns}), No Feedback Utterance Reflections (Table~\ref{tab:likes_nofeedback_utterancereflection}, Table~\ref{tab:dislikes_nofeedback_utterancereflection}), and AI Feedback Utterance Reflections (Table~\ref{tab:likes_aifeedback_utterancereflection}, Table~\ref{tab:dislikes_aifeedback_utterancereflection}). Since session-reflections was the reference condition, the survey did not specifically ask about this condition.

\begin{table*}[htbp]
\centering
\caption{Concerns about CARE's AI feedback system. \textbf{Bolded} values indicate fields where over half of the 25 respondents had concerns.}
\begin{tabular}{|p{12cm}|c|}
\hline
\textbf{Field} & \textbf{Choice Count} \\
\hline
I wish the AI patient could give me feedback from their perspective on how I made them feel & \textbf{17} \\
\hline
I wish CARE's feedback system tracked and referenced how I was progressing over time & \textbf{17} \\
\hline
AI providing unhelpful feedback & \textbf{17} \\
\hline
I wish it gave feedback on the session as a whole & \textbf{16} \\
\hline
I wish its evaluation criteria had more variety each week & \textbf{15} \\
\hline
I wish it gave feedback on my speech, rather than just the written transcript of what I said & \textbf{13} \\
\hline
Other concerns & 6 \\
\hline
\end{tabular}
\label{tab:care_feedback_concerns}
\end{table*}

\begin{table*}[htbp]
\caption{What if anything did you like or find useful about these \textit{utterance-level self-reflection questions}? Note that 15 respondents reported receiving this feature.}
\label{tab:utterance_reflection_likes}
\centering
\begin{tabular}{|p{12cm}|c|}
\hline
\textbf{Field} & \textbf{Choice Count} \\
\hline
I liked getting a second chance to modify my response & 6 \\
\hline
I liked understanding the impact my words had on the AI patients response & 6 \\
\hline
Other: & 2 \\
\hline
\end{tabular}
\label{tab:likes_nofeedback_utterancereflection}
\end{table*}

\begin{table*}[htbp]
\caption{Things you disliked or wish were different about \textit{utterance-level self-reflection questions}? Note that 15 respondents reported receiving this feature, and \textbf{bolded} values indicate fields selected by more than half (8+).}
\label{tab:utterance_reflection_dislikes}
\centering
\begin{tabular}{|p{12cm}|c|}
\hline
\textbf{Field} & \textbf{Choice Count} \\
\hline
I wish I knew why some responses were chosen to reflect upon & \textbf{10} \\
\hline
The standard reflections become repetitive and boring. & \textbf{8} \\
\hline
I wish I could choose which utterances to reflect on & 7 \\
\hline
I prefer to reflect on my overall approach to a practice session & 4 \\
\hline
There were too many of them to answer & 2 \\
\hline
Other dislikes/limitations: & 1 \\
\hline
\end{tabular}
\label{tab:dislikes_nofeedback_utterancereflection}
\end{table*}

\begin{table*}[htbp]
\centering
\caption{What if anything did you like or find useful about these \textit{utterance self-reflection questions accompanying AI feedback}? Note that 10 respondents reported receiving this feature, and \textbf{bolded} values indicate fields selected by more than half (6+).}
\label{tab:utterance_reflection_with_ai_benefits}
\begin{tabular}{|p{12cm}|c|}
\hline
\textbf{Field} & \textbf{Choice Count} \\
\hline
I liked that I could critically engage with the AI feedback, rather than accept it as the gold standard & \textbf{6} \\
\hline
I liked processing my thoughts on the AI feedback by reflecting & 3 \\
\hline
I liked getting a second chance to modify my response & 0 \\
\hline
I liked understanding the impact that one's words could have on an AI patients response & 0 \\
\hline
Other likes/benefits: & 0 \\
\hline
\end{tabular}
\label{tab:likes_aifeedback_utterancereflection}
\end{table*}

\begin{table*}[htbp]
\caption{Things you disliked or wish were different about \textit{utterance-level self-reflection questions accompanying AI feedback}? Note that 10 respondents reported receiving this feature.}
\label{tab:utterance_reflection_with_ai_dislikes}
\centering
\begin{tabular}{|p{12cm}|c|}
\hline
\textbf{Field} & \textbf{Choice Count} \\
\hline
The standard reflections become repetitive and boring. & 4 \\
\hline
I wish I knew why some responses were chosen to reflect upon & 3 \\
\hline
I wish I could choose which utterances to reflect on & 1 \\
\hline
There were too many of them to answer & 1 \\
\hline
Other dislikes/limitations: & 0 \\
\hline
I prefer to reflect on my overall approach to a practice session & 0 \\
\hline
\end{tabular}
\label{tab:dislikes_aifeedback_utterancereflection}
\end{table*}

\begin{table*}[htbp]
\centering
\begin{minipage}{\textwidth}
\centering
\begin{tabular}{lcc|cc}
\toprule
& \multicolumn{2}{c}{Empathy} & \multicolumn{2}{c}{Reflections} \\
& \multicolumn{2}{c}{strong uses} & \multicolumn{2}{c}{strong uses} \\
\cmidrule(lr){2-3} \cmidrule(lr){4-5}
Variable & Estimate & 95\% CI & Estimate & 95\% CI \\
\midrule
Current Performance & 0.043 & -- & 0.043 & -- \\
Intercept (No Feedback, Session Reflection) \textsuperscript{a} & 0.361 & -- & 0.361 & -- \\
No Feedback, Utterance Reflection & -0.071 & (-0.21, +0.06) & -0.071 & (-0.21, +0.06) \\
AI Feedback, Session Reflection & -0.018 & (-0.17, +0.13) & -0.018 & (-0.17, +0.13) \\
AI Feedback, Utterance Reflection & -0.095 & (-0.24, +0.05) & -0.095 & (-0.24, +0.05) \\
\bottomrule
\end{tabular}
\refstepcounter{subtable}\label{subtab:empathy_reflections}
\par\smallskip
{\small \subtablelabel{\thesubtable}\space Treatment Effects on Empathy and Reflections}
\end{minipage}

\vspace{1em}

% Subtable 2: Questions and Suggestions
\begin{minipage}{\textwidth}
\centering
\begin{tabular}{lcc|cc}
\toprule
& \multicolumn{2}{c}{Questions} & \multicolumn{2}{c}{Suggestions} \\
& \multicolumn{2}{c}{strong uses} & \multicolumn{2}{c}{uses needing improvement} \\
\cmidrule(lr){2-3} \cmidrule(lr){4-5}
Variable & Estimate & 95\% CI & Estimate & 95\% CI \\
\midrule
Current Performance & 0.124 & -- & 0.080 & -- \\
Intercept (No Feedback, Session Reflection) \textsuperscript{a} & 0.362 & -- & 0.012 & -- \\
No Feedback, Utterance Reflection & -0.001 & (-0.13, +0.13) & +0.013 & (-0.00, +0.03) \\
AI Feedback, Session Reflection & -0.075 & (-0.25, +0.10) & +0.017$*$ & (+0.00, +0.03) \\
AI Feedback, Utterance Reflection & -0.129$*$ & (-0.25, -0.01) & +0.014 & (-0.01, +0.03) \\
\bottomrule
\end{tabular}
\refstepcounter{subtable}\label{subtab:questions_suggestions}
\par\smallskip
{\small \subtablelabel{\thesubtable}\space Treatment Effects on Questions and Suggestions}
\end{minipage}
\caption{Treatment Effects on Behavioral Performance in Next Practice}
\begin{flushleft}
\footnotesize
\textsuperscript{a} denotes the reference category. \\
Note: $*$ indicates statistical significance. Confidence intervals are based on 2.5\% and 97.5\% quantiles.
\end{flushleft}
\label{tab:treatment_effects_communication}
\end{table*}

% \begin{table}[htbp]
% \centering
% \caption{Treatment Effects on Next Attempt Within 24 Hours (EMEE Analysis)}
% \label{tab:emee_treatment_effects}
% \begin{tabular}{lcc}
% \toprule
% Treatment Condition & Estimate & 95\% CI \\
% \midrule
% Session Reflections Only\textsuperscript{a} & -- & -- \\
% Utterance Reflections Only & -0.028 & (-1.150, 1.100) \\
% AI Feedback, Session Reflections & 0.389 & (-0.261, 1.040) \\
% AI Feedback, Utterance Reflections & 0.260 & (-0.517, 1.040) \\
% \bottomrule
% \end{tabular}
% \begin{flushleft}
% \footnotesize
% \textsuperscript{a} Reference category. \\
% Note: Estimates represent causal excursion effects from micro-randomized trial analysis using EMEE. Outcome is next attempt within 24 hours. Randomization probability = 0.5, numerator probability = 0.5. Confidence intervals are based on 2.5\% and 97.5\% quantiles. None of the treatment effects reached statistical significance (all p > 0.05).
% \end{flushleft}
% \end{table}

\end{document}